\documentclass[%
 reprint,
 superscriptaddress,
bibnotes,
 amsmath,
 amssymb,
 aps,
 prl,
]{revtex4-2}

\usepackage{array}
\usepackage{tabularx}
\usepackage{multirow}
\usepackage{graphicx}% Include figure files
\usepackage{dcolumn}% Align table columns on decimal point
\usepackage{bm}% bold math
\usepackage{bbm}
\usepackage[colorlinks,
			linkcolor=blue,
            anchorcolor=blue,
            citecolor=blue,
            urlcolor=blue,
            breaklinks=true]{hyperref}

\newcommand{\eref}[1]{Eq.~(\ref{#1})}
\newcommand{\fref}[1]{Fig.~\ref{#1}}

\newcommand{\p}{\ensuremath{\partial}}
\newcommand{\im}{\ensuremath{\mathrm{i}}}
\newcommand{\df}{\ensuremath{\mathrm{d}}}
\newcommand{\bra}[1]{\ensuremath{\langle #1|}}
\newcommand{\ket}[1]{\ensuremath{|#1 \rangle}}
\newcommand{\expv}[1]{\ensuremath{\langle #1 \rangle}}

\newcommand{\tr}{\ensuremath{\mathrm{tr}}}

\newcommand{\calE}{\ensuremath{\mathcal{E}}}
\newcommand{\calF}{\ensuremath{\mathcal{F}}}

\newcommand{\calH}{\ensuremath{\mathcal{H}}}
\newcommand{\calI}{\ensuremath{\mathcal{I}}}

\newcommand{\calO}{\ensuremath{\mathcal{O}}}

\newcommand{\calY}{\ensuremath{\mathcal{Y}}}
\newcommand{\ie}{\emph{i.e.}}
\newcommand{\eg}{\emph{e.g.}}

\newcommand{\CR}{Cram\'er-Rao }

\begin{document}

\title{Quantum-inspired Topographic Stereovision}
\author{Fanglin Bao}\email{baofanglin@westlake.edu.cn}
\author{Youfei Xie}
\author{Syed Masood}
\affiliation{Department of Physics, School of Science and Research Center for Industries of the Future, Westlake University, Hangzhou 310030, P. R. China.}
\affiliation{Institute of Natural Sciences, Westlake Institute for
Advanced Study, Hangzhou 310024, P. R. China.}
% \date{\today}

\begin{abstract}
We revisit the conventional triangulation in distant stereovision, when shape rather than distance is the relevant observable. We show through the information-regret analysis that the optimal measurements for absolute distance and relative topography are unexpectedly different and incompatible, exposing an observable-measurement mismatch. To resolve this, we introduce stereo regularization to address stereo anisotropies that violate prevailing emitter-number conservation. Accordingly, we propose a topographic interferometer, which exploits cross-detector correlations to probe topography without measuring the distance profile. Our Fizeau-imaging interferometer turns parallax paths into Mach-Zehnder arms and employs a central path as the local oscillator for balanced homodyne detection, saturating the quantum Fisher information with improved topographic error scaling. This enables topographic stereovision of thermal sources beyond the Rayleigh limit, with feasible experimental demonstrations within existing techniques for remote sensing and astronomy.
\end{abstract}

\maketitle

\begin{figure}[tb]
    \centering
    \scalebox{1}{\includegraphics{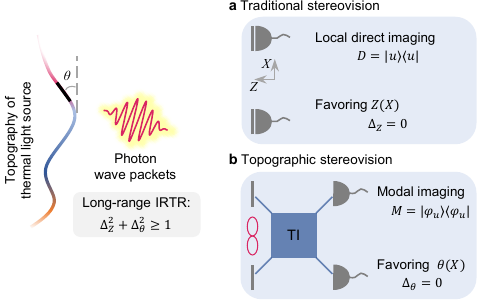}}
    \caption{Quantum-inspired topographic stereovision with topographic interferometer (TI, see \fref{fig:schematic_TI}). In the long-range limit, cross-detector and cross-pixel correlations (b) encode unique topographic information beyond triangulation (a), trading distance precision for improved topographic error scaling. IRTR: information-regret tradeoff relation. $\Delta$: Fisher information regret related to error. $u$: image plane coordinate. $\varphi_u$: modal profile.}
    \label{fig:schematic_QiH}
\end{figure}

%\section{Introduction}
Stereovision by triangulation is a century‑tested principle of remote sensing and astronomy, exemplified in stellar parallax, Cherenkov telescope arrays \footnote{\url{https://www.ctao.org/}}, and emerging heat‑assisted detection and ranging (HADAR) \cite{Bao2023}. Triangulation estimates the target’s absolute distance with a well-known quadratic error scaling, $\delta Z\propto Z^2$. In many stereoscopic settings, however, the more relevant observable is usually the topography, $\theta=\arctan\p_X Z$, as illustrated in \fref{fig:schematic_QiH}. When inferred from the distance measurements, topography has a further-degraded error scaling, particularly manifested in the thermal infrared \cite{Rafael2021} and long-range regimes where significant ghosting effect \cite{Treible_2017_CVPR, *[{}][{. Also see Supple. Mat. II for a brief introduction.}]xu2026} and diffraction hamper stereo matching. This observable-measurement mismatch evokes the long-imagined capability of telescopic vision -- probing distant topography without measuring absolute distance -- and raises a fundamental question: if shape rather than location is desired, is triangulation still the optimal measurement?

The passive nature of astronomical observation precludes the deployment of active quantum techniques such as quantum radar \cite{Barzanjeh2015, Maccone2020} and quantum LiDAR \cite{Reichert2022, Qian2023}. Instead, recent advances in quantum metrology \cite{Tsang2016Aug, Lupo2016} have unveiled quantum-inspired sub-Rayleigh imaging of thermal sources that achieves the quantum \CR bound for unbiased parameter estimation. Demonstrations through modal imaging (\eg, SPADE \cite{Tsang2016Aug}) or interferometry \cite{Lupo2020} (\eg, SLIVER \cite{Nair2016} or HOM interference \cite{Parniak2018}) include super-resolution imaging of point sources with arbitrary brightness \cite{Zhou2019Jan}, 3D location of point sources \cite{Yu2018, Zhou2019}, point number estimation \cite{Lu2018December, Bao2021}, and 2D imaging of extended objects \cite{Tsang2019Jan, Pushkina2021, Lian2026}. When the light source is discretized in the image plane as a set of point emitters, the prevailing paradigm assumes a conserved number of emitters. By contrast, 3D stereovision of extended objects inherently involves multi-detector measurements with angular anisotropy that violates emitter-number conservation.

\begin{figure}[tb]
    \centering
    \scalebox{1}{\includegraphics{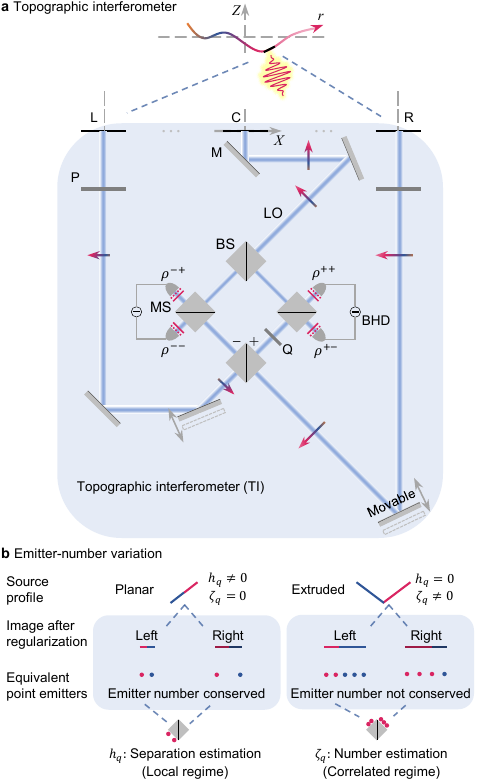}}
    \caption{a. Topographic Fizeau imaging interferometer. A thermal photon emitted by the light source (black line segment) propagates (dashed blue line) to the apparatus and is in a superposition across stereo spatial modes (L: left, R: right), which are fed into the 50:50 beam splitter (BS) for interference. This process resembles a Mach-Zehnder interferometer, in which the contrast between two spatial modes encodes the topography of the thermal light source. We introduce a central spatial mode (C) as the local oscillator (LO) for balanced homodyne detection (BHD) after mode sorters (MS) to fully extract the topographic information. Arrows with varying lengths: virtual images. M: mirror. P: variable phase retarder. Q: quarter-wave phase retarder. Relay lenses are not shown. b. Schematics to illustrate the mechanism of emitter-number variation, which leads to unique topographic information (in $\zeta$) that is accessible only via cross-detector correlations beyond local measurements. Red and blue indicate different brightness. Regularized right images are darker and correspond to fewer point emitters.}
    \label{fig:schematic_TI}
\end{figure}

\begin{figure}[tb]
    \centering
    \scalebox{1}{\includegraphics{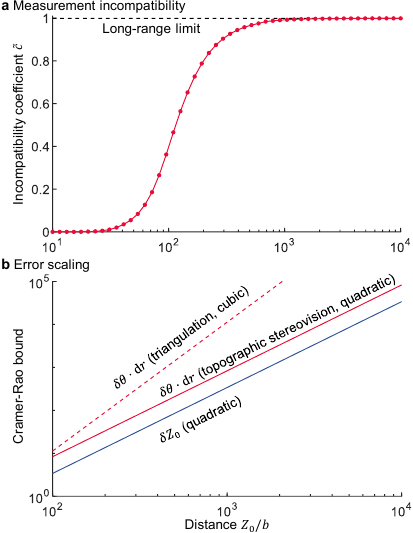}}
    \caption{a. Measurement incompatibility coefficient between distance $Z_0$ and topography $\theta$, as a function of the distance $Z_0$. b. \CR bounds showing the improved topographic error scaling of $\delta\theta$ for fixed photon budgets. Planar object: $\df r=0.1\,\mathrm{m}$, $\theta=\pi/4$, with centroid $(X_0=0,Z_0)$. Baseline $b=1\,\mathrm{m}$. Wavelength: $10\,\mu\mathrm{m}$. Pixel size: $12\,\mu\mathrm{m}$. f-number: 20.}
    \label{fig:IC}
\end{figure}

Here, we introduce stereo regularization to address uncontrolled anisotropy and identify the unique information associated with emitter-number variations that is accessible only through cross-detector correlations (\fref{fig:schematic_TI}). We propose the Fizeau-mode topographic interferometer with optimal positive operator-valued measures (POVMs) to enable topographic stereovision, which serves as a sub-leading effect relative to quantum-inspired imaging. Our information-regret analysis shows that incompatibility in quantum multiparameter estimation of topography and absolute distance emerges at increasing distances (\fref{fig:IC}), necessitating the trading of distance precision for improved topographic error scaling.

Considering a stereovision problem in the $X$-$Z$ plane (\fref{fig:schematic_TI}), without loss of generality. The thermal source has an unknown one-dimensional object profile $(X_r, Z_r)$ with emitted mean photon number distribution $\epsilon^o_r$, where $r$ is the path coordinate along the profile. The interferometer is centered at the origin $O$ with a baseline $b$, focal length $f$, and the apertures are at $X^\mathrm{(L,C,R)}=(- b/2,0,b/2)$ and $Z=0$. Denote $Z_0=\overline{Z_r}$ the mean distance, where the average weight shall be specified later. In the long-range limit, we have $\varepsilon = \max_r\{b,|X_r|,|Z_r - Z_0|\}/Z_0 \ll 1$, and we are interested in the relative topography, $h_r=-{bf(Z_r - Z_0)}/{2Z_0^2}=\calO(\varepsilon)$, where $\overline{h_r}=0$. Note that $h$ and $\p_X h$ are deterministically linked in the continuum limit, we refer to topography as either relative distance or distance gradient throughout the paper.

\textbf{Quantum description--} The incoherent thermal source can be discretized into $Q$ tiny segments with centroid $r_q=(X_q, Z_q)$, tilt angle $\theta_q$, and infinitesimal length $\df r$, $q=1,2,\cdots,Q$. In the long-range and weak-source limits where the received mean photon number $\epsilon \sim \max|X(r)|^2/Z_0^2 =\varepsilon^2 \ll 1 $, the optical field with given polarization and frequency $\omega$ entering the interferometer has an expansion (see Supple. Mat. I),
\begin{equation}\label{eq:rhoImgExp}
    \rho = (1-\epsilon)\rho_0 + \epsilon\rho_1 + \calO(\epsilon^2).
\end{equation}
Let $u$ denote the image-plane coordinate, $\ket{0}=\otimes_u\ket{0}_u$ the joint vacuum state, and $\ket{u}=a^\dagger_u\ket{0}$ the state with one photon in the spatial mode $u$. We have $\rho_0 = \ket{0}\bra{0}$, and $\rho_1 = \sum_q I_q\ket{\Psi_q}\bra{\Psi_q}$, where
\begin{equation}\label{eq:sps}
    \ket{\Psi_q} = \sqrt{\frac{\eta_q^\mathrm{L}}{\eta_q^\mathrm{C}}}e^{\im \phi_q}\ket{\psi^\mathrm{L}_q} + \ket{\psi^\mathrm{C}_q} + \sqrt{\frac{\eta_q^\mathrm{R}}{\eta_q^\mathrm{C}}}e^{-\im \phi_q}\ket{\psi^\mathrm{R}_q}.
\end{equation}
Here, $\ket{\psi^i_q} = \sum_{u\in i} S_{uq}/\sqrt{\eta^i_q}\,\ket{u}$ is the single-photon state of the $i$-th detector mode ($i=\mathrm{L, C, R}$), $S_{uq}$ is the scattering matrix describing the linear but dissipative propagation from source mode $q$ to image mode $u$, and $\eta_q^i=\sum_{u\in i}|S_{uq}|^2$ is each detector's efficiency with $\eta_q=\sum_i\eta^i_q\le 1$. Typically, $\eta^i_q \propto A^i\cdot\mathrm{d}x^i/\mathrm{d}r$ encodes the attenuation $A^i$ and the image length $\mathrm{d}x^i$ of the source segment. The unnormalized intensity distribution is $I_q=\epsilon^o_q\eta_q^\mathrm{C}/\epsilon = \epsilon^o_q\eta_q^\mathrm{C}/\sum_q \epsilon^o_q \eta_q$. The phase contrast among stereo detectors is $\phi_q=\omega bX_q/2cZ_q$, where $c$ is the speed of light.

The optical field acquires propagated coherence in \eref{eq:sps}, and in the weak-source limit the thermal state $\rho$ [second-order coherence $g^2(0)=2$] becomes a single-photon state $\rho_1$ [$g^2(0)=0$] by neglecting multi-photon events, \eg, through post selection or by restricting the number of temporal modes (measurements) $T$ to be $1/\epsilon\ll T\ll1/\epsilon^2$. If we focus on the stereo arms and denote $\ket{\psi^\mathrm{L}_q}=\ket{10}$, $\ket{\psi^\mathrm{R}_q}=\ket{01}$ (ignoring coefficient difference), the single-photon wave packet is in a superposition across two detector modes, which can be formally described as a Bell state of maximal qumode entanglement \cite{Jia2025}, implying that local measurement on each detector such as triangulation might be non-optimal.

\textbf{Stereo regularization--} In adopting quantum-inspired imaging for stereovision, two unique features arise: anisotropic projection and anisotropic scattering. Firstly, according to ray optics, the projected image length of the line segment, $\df x^i = \frac{f\cdot\df r}{Z}[\cos\theta-\sin\theta\cdot \frac{X-X^i}{Z}]$, varies across detectors. When images are discretized into virtual point emitters, one detector sees that the emitter count in other detectors varies (\fref{fig:schematic_TI}b), in stark contrast to the prevailing paradigm of emitter-number conservation. Secondly, detectors with different perspectives observe anisotropic scattering from extended objects, depending on the unknown environment and the Bidirectional Reflectance Distribution Function (BRDF). We emphasize that in the long-range limit, the scattering anisotropy can be modeled as a total-intensity correction with coefficient $B$, consistent with traditional stereovision, which has proven successful so far when using normalized pattern matching instead of absolute-intensity search. These two anisotropies lead to $\sqrt{\eta^\mathrm{L}/\eta^\mathrm{C}}=\sqrt{A^\mathrm{L}\df x^\mathrm{L}/A^\mathrm{C}\df x^\mathrm{C}}\cdot(1-B\frac{b}{2Z_0})\approx (1-\beta)a_1+\calO(\varepsilon^2)$, $\sqrt{\eta^\mathrm{R}/\eta^\mathrm{C}}=\sqrt{A^\mathrm{R}\df x^\mathrm{R}/A^\mathrm{C}\df x^\mathrm{C}}\cdot(1+B\frac{b}{2Z_0})\approx (1+\beta)a_2+\calO(\varepsilon^2)$, where $\beta =\frac{b}{2Z_0} \frac{\p_X Z}{2} + B\frac{b}{2Z_0}=\calO(\varepsilon)$, $a_1=\sqrt{A^\mathrm{L}/A^\mathrm{C}}$, and $a_2=\sqrt{A^\mathrm{R}/A^\mathrm{C}}$. To mitigate the uncontrolled scattering, we introduce a stereo regularization that actively sets the attenuation to minimize the integrated absolute difference between the left and right images,
\begin{equation}
    \begin{split}
        &\mathrm{argmin}_{a_i} |\tr(\rho^\mathrm{R}-\rho^\mathrm{L})|\\
        =\ &\mathrm{argmin}_{a_i} |(a_2^2-a_1^2)+2\bar{\beta}(a_2^2+a_1^2)|+\calO(\varepsilon^2),
    \end{split}
\end{equation}
where $\rho^j=\sum_qI_q\eta^j_q/\eta^\mathrm{C}_q\ket{\psi^j_q}\bra{\psi^j_q}$ is the unnormalized conditioned density operator, $i=\mathrm{argmax}_j\tr\rho^j\in\{\mathrm{L,R}\}$, and we define the average as $\bar{\beta}:=\sum_q I_q \beta_q/\sum_q I_q$ throughout the paper. The regularization condition is
\begin{equation}
    a_2=(1-\bar{\beta})a, \quad a_1=(1+\bar{\beta})a, \quad a=\sqrt{a_1a_2}.
\end{equation}
Note that stereo regularization also applies when the scattering anisotropy has been calibrated or is negligible ($B=0$).

\textbf{Interferometer with optimal POVMs--} Now we operate the interferometer to illustrate topographic stereovision. Two calibrated movable mirrors move jointly so that their lateral beam shifts have the same magnitude but opposite directions. Denote $x_q=x_q^\mathrm{C}$ the central image position, and let $\tilde{x}$ denote the shifted stereo image position, $\tilde{x}^\mathrm{L}_q=x^\mathrm{L}_q + \Omega$, $\tilde{x}^\mathrm{R}_q=x^\mathrm{R}_q - \Omega$, $\Omega$ being the untracked lateral translation. We roughly align the image centers by blocking each arm and observing the image at the symmetric port $\ket{+}$ of the bottom beam splitter, thereby pushing the remaining disparity between the stereo detectors, $h_q = \tilde{x}^\mathrm{R}_q - x_q = x_q - \tilde{x}^\mathrm{L}_q$, into the sub-Rayleigh regime, $|\overline{h_q}|\ll \sigma_0$, where $\sigma_0$ is the width of the point-spread function (PSF). Note that the remaining disparity $h_q$ captures the rescaled relative topography (which is an unknown linear transformation of $h_r$),
\begin{equation}
    h_q = -\frac{bf(Z_q-Z_\delta)}{2Z_\delta^2}=\calO(\varepsilon),
\end{equation}
where $Z_\delta = bf/2\Omega = Z_0+\Delta H$, and $\Delta H=2\overline{h_q} \cdot Z_0^2/bf\ll Z_0$. Nonzero $\overline{h_q}$ implies that the interferometer is not gazing at the mean absolute distance $Z_0$ but instead at $Z_\delta$. We emphasize that the above alignment is intended to roughly establish a gazing reference $Z_\delta$ for later series expansion, not to calibrate the mean distance $Z_0$ or measure the distance profile $Z_q$. The observable of the topographic interferometer is the rescaled relative topography $h_q$ reflecting the shape \footnote{This is an interferometric version of the classic experiments by Bond and De La Rue in the 1860s, which exploited lunar libration to reveal its mountains and craters in striking depth, even though the Earth–moon distance itself was not measured.}. It is possible to recover the absolute distance by precisely tracking the coarse disparity $\Omega$, but it again suffers from quadratic error scaling. Now, we expand functions of $\tilde{x}_q$ on the beam splitter in terms of $x_q$ to first order. That is, $\ket{\psi(\tilde{x})}^\mathrm{R}_q = \ket{\psi(x)}^\mathrm{R}_q + \ket{\psi'(x)}^\mathrm{R}_q h_q$, and $\ket{\psi(\tilde{x})}^\mathrm{L}_q = \ket{\psi(x)}^\mathrm{L}_q - \ket{\psi'(x)}^\mathrm{L}_q h_q$. Putting together, we have the optical field with explicit topographic dependence,
\begin{equation}\label{eq:stereoRho}
    \begin{split}
        \ket{\Psi_q} &= a[ e^{\im \phi_q}(1-\zeta_q)(\ket{\psi_q^\mathrm{L}} - \ket{\psi_q^\mathrm{L}}' h_q) + \frac{1}{a}\ket{\psi_q^\mathrm{C}} \\
        & + e^{-\im \phi_q}(1+\zeta_q)(\ket{\psi_q^\mathrm{R}} + \ket{\psi_q^\mathrm{R}}' h_q) ],
    \end{split}
\end{equation}
where $\zeta_q=\beta_q-\bar{\beta}=\frac{1}{2}[\p_xh-{\overline{\p_xh}}]$, $\overline{\zeta_q}=0$, and $\p_X=-f/Z_\delta\cdot\p_x$. Note that topographic information in $h$ is the contribution as if the object is a set of point emitters with fixed counts, while the topographic information in $\zeta$ is unique in stereovision that captures fine emitter-number variations, see \fref{fig:schematic_TI}b.

Here, we sketch the main results and highlight the significance, with details given in Supple. Mat. III-V. For a set of POVMs $E_n$, $\sum_nE_n=\mathbb{I}$, the four interferometer output probabilities can be described by two complex numbers,
\begin{equation}
    p_n =\tr(\rho E_n)= \frac{1}{4}\int I_x\expv{E_n}_x\df x \pm \frac{1}{2}\Delta p_n,
\end{equation}
in the continuum ($\df x\to 0$) and strong-oscillator ($a\to 0$) limits, where $\expv{E_n}_x=\bra{\psi_x}E_n\ket{\psi_x}$. The left/right detections measure the real and imaginary parts, and the top/bottom detections correspond to plus/minus signs, respectively. The balanced homodyne detection (BHD) signals are
\begin{equation}
    \Delta p_n = a\int I_x e^{\im\phi_x}\left[ \p_x(h_x\expv{E_n}_x) - \overline{\p_xh_x}\expv{E_n}_x \right]\df x.
\end{equation}
We define the central image's moments,
\begin{equation}
    \calI_{j} = a^2\int I_x (\frac{x-\bar{x}}{\sigma_0})^j\df x = \calO(\varepsilon^j),
\end{equation}
where $\bar{x}=\int I_x x\df x/\int I_x\df x$ is the image center. We then employ the following expansions that converge in the long-range limit,
\begin{gather}
    \expv{E_n}_x = \sum_{j=0}^\infty \calE_{nj} (\frac{x-\bar{x}}{\sigma_0})^j,\nonumber \\
    e^{\im \phi_x} = \sum_{k=0}^\infty \calF_k (\frac{x-\bar{x}}{\sigma_0})^k,\\
    h_x = \sum_{m=0}^\infty \calH_m (\frac{x-\bar{x}}{\sigma_0})^m.\nonumber
\end{gather}
Here, $\calE_{nj}=\sigma_0^j[\p_x^j\expv{E_n}_x]_{\bar{x}}/j!$, $\calF_k = (-\im b\omega\sigma_0/2cf )^k/k!$, and $\calH=\calO(\varepsilon)$. We find that
\begin{equation}
    p_n = \frac{1}{4a^2}\sum^\infty_{j=0} \calE_{nj}\calI_j \pm \frac{1}{2a}\sum^\infty_{j=0} \calE_{nj}\mathcal{Z}_j,
\end{equation}
where $\mathcal{Z}_j=\sum_m \mathcal{T}_{jm} \calH_m = \calO(\varepsilon^j)$ is the topographic moment with kernel
\begin{equation}
        \mathcal{T}_{jm} = \sum_{k}\calF_k \left\{ \frac{j+m}{\sigma_0} \cdot \calI_{j+k+m-1} - \frac{m}{\sigma_0}\cdot \calI_{j+k}\frac{\calI_{m-1}}{\calI_{0}}  \right\} .
\end{equation}
The optimal POVMs for estimating the topographic moments $\mathcal{Z}_{2j}$ and $\mathcal{Z}_{2j+1}$ (assuming a Gaussian aperture) saturate the quantum Fisher information
\begin{equation}
    K_{\mathcal{Z}} = \max_{\{E_n\}}\sum_n(\p_\mathcal{Z} p_n)^2/p_n,
\end{equation}
and are given by
\begin{equation}
    E_n\to M^{\pm}_{j,j+1}(u)=\ket{\varphi^{\pm}_{j,j+1}(u)}\bra{\varphi^{\pm}_{j,j+1}(u)},
\end{equation}
where $\ket{\varphi^{\pm}_{j,j+1}(u)}=[\ket{\varphi^j_{\bar{x}}(u)}\pm \ket{\varphi^{j+1}_{\bar{x}}(u)}]/\sqrt{2}$, and $\varphi^j$ is the $j$-th Hermite-Gaussian mode, which renders $\calE_{nk}=0, \forall k<2j$. From the BHD signals $\Delta p_n$, topography $\calH_m$ can be reconstructed by the following linear system,
\begin{equation}
    \Delta p_n = \frac{1}{a}\sum_{m=0}^\infty \calY_{nm}\calH_m,
\end{equation}
where $\calY_{nm} = \sum_{j}\calE_{nj} \mathcal{T}_{jm}$. Note that image moments $\calI$ can be pre-calibrated by summing the four outputs, $p^\mathrm{tot}_n = \sum_j\calE_{nj}\calI_j/a^2$, and coefficients $\calE_{nj}$, $\calF_k$, and hence $\calY_{nm}$, can be completely determined with known interferometer configurations.

We underscore the key role of the central arm by \eref{eq:stereoRho}, where discrete $h$ (relative distance) and $\zeta$ (relative angle) are independent. In the short-baseline limit ($\phi_q\equiv 0$), the optical field after the bottom beam splitter is (see Eq.~18 of the Supple. Mat.), $\ket{\Psi}=\sqrt{2}a(\ket{+}+\zeta\ket{-}+h\ket{-}'+\ket{\psi^\mathrm{C}}/\sqrt{2}a)$, where we can see the topographic information lives only in the antisymmetric mode $\ket{-}$, and so will be the optimal POVMs, $E_n\ket{+}=E_n\ket{\psi^\mathrm{C}}=0$. Consequently, for a planar object (\fref{fig:schematic_TI}b left, $\zeta=0$), we have $p_n=\sum_q 2Ia^2 h^2\bra{-'}E_n\ket{-'}=\calO(\varepsilon^2)$, $\p_h p_n =4Ia^2 h\bra{-'}E_n\ket{-'}=\calO(\varepsilon)$, and for a single segment, $K_h = 8Ia^2\max_{\{E_n\}}\sum_n\bra{-'}E_n\ket{-'} =2Ia^2/\sigma_0^2$, coinciding with the quantum Fisher information for point-separation estimation ($2h$ is the separation, $N=2Ia^2$ is the total photon number) \cite{Tsang2016Aug}. On the contrary, traditional stereovision relies on local direct imaging, $D_u=\ket{u}\bra{u}$, with $p^i_u=\tr(\rho^i D_u)=\calO(1)$. The derivative is in the stereo contrast, $\p_h p_u\propto \p_h\tr[(\rho^\mathrm{R}-\rho^\mathrm{L})D_u]=\calO(\varepsilon)$, due to the fact that $\bar{h}_r=0$, and its classical Fisher information is $J^{\mathrm{D}}_h = \calO(\varepsilon^2)\ll K_h$. When $\phi_q\neq 0$ for long baselines, however, topographic information is distributed across both symmetric and antisymmetric modes and buried in the phase variation. Nevertheless, we observed that topography is encoded mainly in the real part of the optical field (after the quarter-wave phase retarder). The optimal measurement is therefore the quadrature POVM with zero phase, implemented via BHD, with the central mode as the local oscillator. The final classical Fisher information of the interferometer saturates the quantum Fisher information (single segment),
\begin{equation}
    \frac{J^{\mathrm{TI}}_h}{K_h} = \left[ \frac{\sin^2\phi}{(4a^2\cos^2\phi+1)} + \frac{\cos^2\phi}{(4a^2\sin^2\phi+1)} \right]\to 1,
\end{equation}
in the strong-oscillator limit, for fixed photon budgets on side stereo detectors. This limit can be achieved, \eg, by increasing the bandwidth of the central arm or attenuating the stereo arms. Note that even with a weak oscillator, $a\sim1$, the interferometer can still work near optimal, $J^{\mathrm{TI}}_h/K_h=\calO(1)$.
% BHD directly on the stereo arms is not viable, as source-dependent phases cannot be eliminated by a single quadrature phase.

We argue that the topographic information in $\zeta$ (\fref{fig:schematic_TI}b right, $h=0$) associated with emitter-number variation can only be retrieved by cross-detector correlations, not even by adopting SPADE in each stereo detector. Note that $p_n=\sum_q 2Ia^2\zeta^2\bra{-}E_n\ket{-}=\calO(\varepsilon^2)$, $\p_\zeta p_n=4Ia^2\zeta\bra{-}E_n\ket{-}=\calO(\varepsilon)$, and $K_{\zeta} = \calO(1)$. However, to keep $\zeta$-dependence in SPADE (see Eq.~13 of the Supple. Mat.), we have $\bra{\psi^i}E_\mathrm{spade}\ket{\psi^i}_{\bar{x}} \neq 0$, leading to $p^i=\tr(\rho^i E_\mathrm{spade})=\calO(1)$. Since $\bar{\zeta}_q=0$, $\p_\zeta p^i=\calO(\varepsilon)$, the classical Fisher information by local SPADE is vanishingly small, $J_{\zeta} = \calO(\varepsilon^2)\ll K_{\zeta}$, reflecting the unique advantage of our interferometer. Also note that local SPADE does not require cross-detector coherence and enables very long baselines for stellar observations, at the cost of the $\zeta$-dependent topographic information.
% Furthermore, even with SPADE super-resolution imaging (through $\delta$) in each stereo detector, the ghosting effect still hampers accurate stereo matching \cite{Treible_2017_CVPR, *xu2026}, reflecting the unique advantage of our interferometer.

\textbf{Scaling law and incompatible measurements--} The scaling of quantum Fisher information is $K=\calO(\varepsilon^{2k-2})$ for $\mathcal{Z}_{2k}$, $K=\calO(\varepsilon^{2k})$ for $\mathcal{Z}_{2k+1}$, and the classical Fisher information by triangulation is $J=\calO(\varepsilon^{2k-2})$ for $\mathcal{Z}_{k}$ \cite{Zhou2019Jan}. Therefore, the ranging error in terms of the $k$-th order topographic moment has an improvement
\begin{equation}
    \delta \mathcal{Z}_k^{\mathrm{TI}}/\delta \mathcal{Z}_k^{\mathrm{D}} \propto (b/Z)^{-\lfloor k/2 \rfloor},
\end{equation}
for either fixed exposure time or photon budgets, where $\lfloor \cdot \rfloor$ is the floor symbol. The corresponding optimal measurement for topography, \ie, interference followed by modal imaging, is incompatible with triangulation for the absolute distance. Note that decreasing the absolute distance destroys the cross-detector and cross-pixel coherence; hence, in the short-range limit, incompatibility disappears, and the optimal measurement reduces to traditional triangulation.

\fref{fig:IC} demonstrates the incompatibility evolution and the improved topographic error scaling (see Supple. Mat. VI for details), for a planar object (\eg, space debris) with uniform brightness.
Here, the information-regret tradeoff relation (IRTR) in quantum multiparameter estimation of distance $Z$ and tilt angle $\theta$ is \cite{Lu2021},
\begin{equation}
    \Delta_Z^2 + \Delta_\theta^2 + 2\sqrt{1-\tilde{c}^2}\Delta_Z\Delta_\theta\ge \tilde{c}^2,
\end{equation}
where $\Delta_\mu=\sqrt{(K_{\mu}-J_{\mu})/K_{\mu}}$ is the Fisher information regret related to measurement errors for parameter $\mu=\{Z, \theta\}$. 
In the long-range limit, we have $\Delta_Z^2 + \Delta_\theta^2 \ge 1$. While triangulation strives for absolute distance ($\Delta_Z=0$), topographic stereovision offers an alternative: less distance precision, more shape perception ($\Delta_\theta=0$).

Our work enables topographic stereovision of thermal sources beyond the Rayleigh limit. The topographic interferometer can be experimentally implemented with existing techniques (see Supple. Mat. VII for examples) and generalized to large photon numbers \cite{Lupo2016} and large baselines \cite{Padilla2026, Stas2026}. A general BRDF in the short range may require calibration and deserves further study. However, we believe our topographic stereovision opens new opportunities, \eg, for establishing a quantum-inspired framework of HADAR topography, with applications in remote sensing, astronomy, and non-invasive topographic microscopy.

This work was supported by the National Natural Science Foundation of China. We thank X.-M. Lu for helpful discussions.

\bibliography{QH}
\end{document}